\documentclass[letter]{aa} 
\usepackage{graphicx}
\usepackage{txfonts}
\usepackage{natbib}
\bibpunct{(}{)}{;}{a}{}{,}
\newcommand{\doceCO}{\mbox{$^{12}$CO}}

\newcommand{\jdu}{\mbox{$J$=2$-$1}}
\newcommand{\juc}{\mbox{$J$=1$-$0}}

\newcommand{\kms}{\mbox{km\,s$^{-1}$}}

\newcommand{\ms}{\mbox{$M_{\mbox{\sun}}$}}

\newcommand{\lsim}{\raisebox{-.4ex}{$\stackrel{\sf <}{\scriptstyle\sf \sim}$}}

\begin{document}
   \title{The nebula around the post-AGB star 89\,Her \thanks{Based on 
 observations carried out with the IRAM Plateau de Bure Interferometer,
 as well as on observations of the Belgian Guaranteed time on
 VISA (ESO). 
IRAM is supported by INSU/CNRS (France), MPG (Germany) and IGN
 (Spain).}
}

   \author{
          V. Bujarrabal\inst{1}
          \and
          H. Van Winckel\inst{2}
	  \and
          R. Neri\inst{3}
          \and
          J. Alcolea\inst{4}
          \and
          A. Castro-Carrizo\inst{3}
          \and
	  P. Deroo\inst{2}
          }


   \institute{             Observatorio Astron\'omico Nacional (OAN-IGN),
              Apartado 112, E-28803 Alcal\'a de Henares, Spain\\
              \email{v.bujarrabal@oan.es}
              \and
              Instituut voor Sterrenkunde, 
              KU Leuven, Celestijnenlaan 200B, 3001 Leuven, Belgium\\
	      \email{Pieter.Deroo@ster.kuleuven.ac.be, 
Hans.VanWinckel@ster.kuleuven.be}
              \and
              Institut de Radio Astronomie Millim\'etrique (IRAM),
              300 Rue de la Piscine, F-38406 St.-Martin d'H\`eres, France\\
              \email{ccarrizo@iram.fr, neri@iram.fr}
              \and
              Observatorio Astron\'omico Nacional (OAN-IGN), 
              C/ Alfonso XII 3, E-28014 Madrid, Spain\\
              \email{j.alcolea[at]oan.es}
            }

   \date{Received ; accepted }

  \abstract
   {}
   {We aim to study the structure of the nebula around the post-AGB,
  binary star 89\,Her. The presence of a rotating disk around this star had
  been proposed but not been yet confirmed by observations.}
   {We present high-resolution PdBI maps of CO $J$=2--1 and 1--0. 
Properties of the nebula are directly derived from the data and
  model fitting. We also present  N-band interferometric data on the
  extent of the hot dust emission, obtained with the VLTI.}
   {Two nebular components are found: (a) an extended hour-glass-like
  structure, with expansion velocities of $\sim$ 7 \kms\ and a total mass
  $\sim$ 3 10$^{-3}$ \ms, and (b) an unresolved very compact
  component, smaller than $\sim$ 0\farcs 4 and with a low total
  velocity dispersion of $\sim$ 5 \kms. We cannot determine the velocity
  field in the compact component, 
  but we argue that it can hardly be in expansion, since
  this would require too recent and too sudden an ejection of mass. On the
  other hand, assuming that this component is a keplerian disk, we
  derive disk properties that are compatible with expectations for such
  a structure; in particular, the size of the rotating gas disk should
  be very similar to the extent of the hot dust component  from 
  our VLTI data. Assuming that the equator of the extended nebula 
  coincides with the binary orbital plane, we provide new 
  results on the companion star mass and orbit. 
}  
   {}
   \maketitle
%

\section{Introduction}

Planetary and protoplanetary nebulae (PNe, PPNe) very often show
axisymmetric shapes and fast axial expansion, which are thought to be
due to shock interaction between the very collimated post-AGB jets and
the slow and isotropic AGB wind. Theoretical calculations show
that accretion from rotating disks onto the post-AGB star or a
companion can explain the axial jets \citep[see][etc]{soker02,frankb04}
and provide the high energy and momentum required to explain the PPN
dynamics \citep{bujetal01}.  On the other hand, the presence of a large
amount of orbiting material has been invoked to explain some remarkable
observational properties: 1) the peculiar abundances sometimes found in
the atmospheres of post-AGB stars, where refractory elements are
deficient, presumably because of reaccretion of material in which
grains have been efficiently formed 
(removing such elements from the gas) 
and expelled \citep{vanwinckel03}; and 2) the detection of a NIR
excess due to dust at temperatures close to that of sublimation, which
should correspond to stable structures (i.e. in keplerian rotation)
close to the star. Such SEDs are rather common among post-AGB
stars and, in a review paper of \citet{deruyter06}, fifty one objects are
listed.  These features are systematically found to be associated to
binarity, which would explain the excess of angular momentum needed to
form keplerian disks from previously ejected gas.

   \begin{figure*}
\vspace{-1.5cm}
\hspace{-0.5cm}
\rotatebox{270}{
\resizebox{14cm}{!}{ \includegraphics{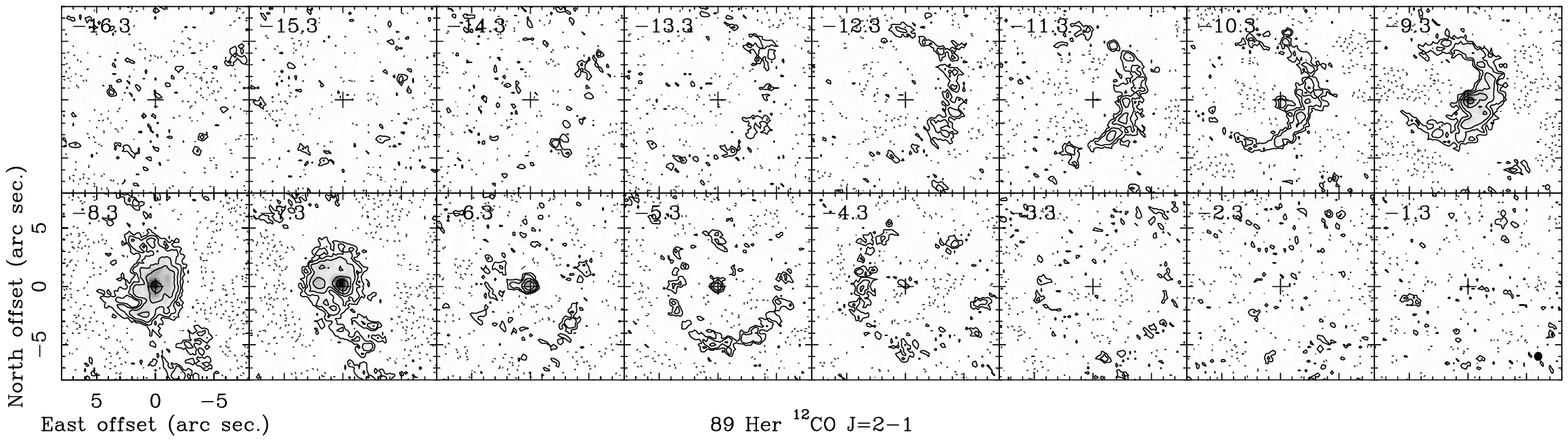}
}}

\vspace{-9cm}
\hspace{-0.5cm}
\rotatebox{270}{
\resizebox{14cm}{!}{ \includegraphics{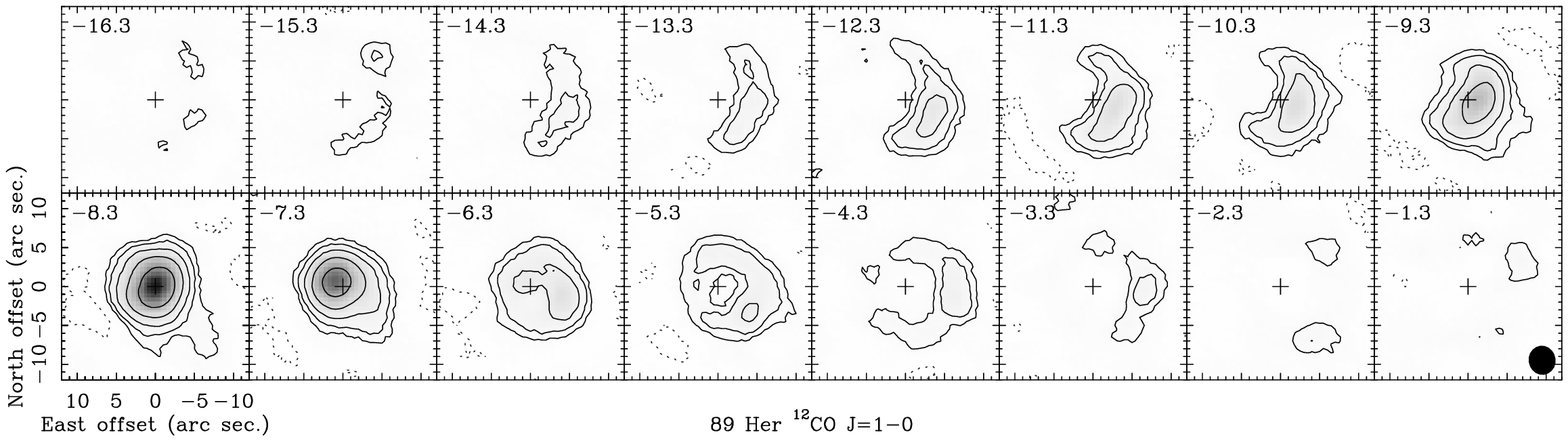}
}}

\vspace{-7.5cm}
   \caption{Channel maps of the \doceCO\ \jdu\ (upper panels) and \juc\
            lines (lower panels) from 89\,Her. The first contours are,
            respectively, 15 and 30 mJy; the contours are separated by
            a factor 2 and the negative contours (at -15 and -30 mJy,
            respectively) are indicated by dashed lines. The LSR 
            velocity in \kms\ for the center of each channel is
            indicated in the upper left corner. 
The J2000 coordinates for the reference
            position, the cross in the maps, are R.A.\ 15:55:25.19 and
            Dec.\  +26:03:00.0. The black ellipses in the last panels 
indicate the beam half-intensity sizes.}
          
              \label{maps}%
    \end{figure*}

The existence of rotating disks around post-AGB stars is 
therefore a basic question to understand the post-AGB
evolution, in particular, the 
shape and dynamics of PPNe.  

Disks or tori of molecular gas around post-AGB stars are commonly
detected as the central part of PPNe, although in general they are not
observed to rotate, but to be in expansion, like the rest of the nebula.
It has been suggested \citep[e.g.][]{fong06} that wide oblate
structures found around a few AGB stars could be in rotation, but the
evidence is very controversial \citep[][etc]{josselin00,hirano04}.
Direct detection of a keplerian velocity field has been
obtained for only one PPN till now: the Red Rectangle
\citep{bujetal05}, a well-known PPN surrounding a binary star.


89\,Her (HR\,6685) was first studied by 
\citet{bidelman51}, who noted the presence of a presumably massive
supergiant at high Galactic latitude. There is now, however, general
consensus that 89\,Her is a low-mass object in its post-AGB phase of
evolution. It is one of the few post-AGB stars with a significant
Hipparcos parallax, placing it at 1$^{2.3}_{0.6}$ kpc from the Sun.
Interestingly, 89\,Her is a binary system that shows both a conspicuous
NIR excess, due to hot dust \citep{waters93,deruyter06}, and a far-IR
flux indicating the presence of large grains \citep{shenton95}. Its CO
lines show a peculiarly sharp profile very similar to the one detected in
the Red Rectangle
\citep{bujetal01}. Previous interferometric
maps of the CO emission from 89\,Her 
\citep{alco95,fong06} 
have not yielded  
conclusive results, due to the small size and low intensity of the
source. A roughly rounded structure was found, showing a central
core plus an extended clumpy halo, with no clear velocity
pattern. 

\section{Observations}

We report on high-resolution interferometry with
the IRAM Plateau de Bure Interferometer of the $^{12}$CO(1--0) and
$^{12}$CO(2--1) lines and continuum emission from 89\,Her. Observations
were carried out under excellent atmospheric conditions (pwv =
0.5-2\,mm, seeing $\simeq$ 0\farcs 1--0\farcs 4) with the six-element array in
the D configuration (November 11, 2005) and in the newly extended A
configuration (January 19 and 25, 2006). The cross-correlator was
set up to cover the lines with 20 and 40\,MHz bands each and the
continuum emission with two 320\,MHz bands. The flux calibration was
referred to MWC349 and 3C273, while phase and amplitude calibrations
were done using 1751+288.

Data were calibrated in the antenna-based manner using the
GILDAS\footnote{The GILDAS software package is available for download
on the IRAM web pages.} software package. Velocity maps were produced
with natural weights at 2.6 mm in order to analyze the extended
molecular distribution and by applying an inverse Gaussian taper at
1.3 mm to analyze in detail the compact emission in the immediate
surroundings of 89\,Her. Cleaned and primary beam-corrected maps were
obtained with synthesized beams of 3\farcs 60$\times$3\farcs 32 at
position angle 
$-97^\circ$ and 0\farcs 68$\times$0\farcs 60 at $-97^\circ$, 
respectively at 2.6 mm and 1.3 mm.

The line emission was resampled to an effective velocity resolution of
1 \kms\ yielding a one-sigma sensitivity level of 
$\sim9.3$\,mJy/beam for the $^{12}$CO(1--0) and of
$\sim15.3$\,mJy/beam for the $^{12}$CO(2--1) line. The continuum
sensitivities were estimated from the line-free emission regions of
the bands to be $\sim0.3$\,mJy/beam at 2.6\,mm and $\sim0.5$\,mJy/beam at
1.3\,mm. Continuum emission was well-detected at 2.6\,mm and 1.3\,mm
but remained unresolved, 
yielding flux densities of 2.7\,mJy and 9.2\,mJy, respectively, 
compatible with thermal dust emission. The size of the continuum source
is smaller than 0\farcs 3.

\subsection{VLTI data}

We also performed N-band interferometric observations using the
two-beam VLTI/MIDI instrument. Three set-ups were used in April 2006
during Belgian guaranteed time on the auxiliary telescopes (VISA). For
one of the observations, the baseline of 16 m (projected angle = 78
degrees) was used, while the other two were conducted on a larger
baseline of 32 m with similar projected angles (80, 77 degrees). The
visibilities were calibrated using observations of unresolved
calibration stars (HD\,168454 and HD\,123139), which were observed in
strict concatenation with the science target. These observations were
performed in the context of a large program to systematically study the
circumstellar discs around binary post-AGB stars.
With the 1.8m auxiliary telescopes, the sensitivity
of MIDI allowed fringe detection on a correlated flux larger than 20 Jy
at 12 $\mu$m. 
For the 
details on the data reduction, we refer the reader to \cite{deroo06}. 

\section{Nebula structure: a rotating disk around 89\,Her?}

As seen in our maps (Fig.\ \ref{maps}), the nebula around 89\,Her
consists of two 
very different features, a quite extended one and a very compact
central clump.  The shape of the extended component strongly suggests
an expanding hour-glass-like structure, whose axis is slightly inclined
with respect to the line of sight. The extended component must have a
maximum expansion velocity $\sim$ 6 - 7 \kms\ (the velocity probably
increases with the distance to the star, in view of the variations in 
the image size with the velocity and as is confirmed by the model
calculations explained below). For a distance of 1 kpc (Sect.\ 1), the
typical total size of this extended component is estimated to be $\sim$
1.5 10$^{17}$ cm.

We do not detect the extent of the compact component, which must be
smaller than $\sim$ 0\farcs 4, equivalent to a diameter smaller than
$\sim$ 5 10$^{15}$ cm. This size limit is comparable to that of the
dust emitting at 1mm wavelength, and probably both emissions come from
the same component. Note that interferometric maps of the
optically-thin dust emission at this wavelength strongly select the
massive compact components. The total velocity dispersion is $\sim$ 5
\kms, with no significant velocity pattern. A slight velocity gradient
in the north-east south-west direction could be seen in our $J$=2--1
maps of the central clump, which could be consistent with the trend
found for the extended component. But note that contamination by the
emission of the wide component may be relevant in this respect, as it
happens in the $J$=1--0 maps, where the extended component emission
always dominates the brightness distribution due to the relatively wide
beam.
%
An inner E-W elongation also appears in the low-resolution maps by
\citet{fong06}, but our data clearly show that it is due to emission
from the hour-glass-like structure.

We have tried to reproduce our data by synthetic maps from an axially
symmetric model nebula. The model description and assumptions can be
found elsewhere \citep[e.g.][]{bujetal98,ccarrizo02}. We considered an 
hour-glass-like extended structure, with very thin walls, 
plus an equatorial compact disk. 

\begin{figure}

\vspace{-0.2cm}
\hspace{0.7cm}
\rotatebox{270}{
\resizebox{6.2cm}{!}{ \includegraphics{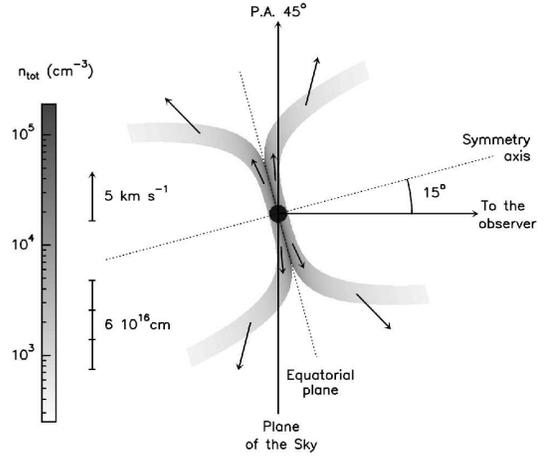}}}

   \caption{Geometry and 
distribution of density and velocity in the model nebula. Note
            that we only obtained limits for the size and physical 
conditions of the central, compact component, represented by the black
circle.} 
          
              \label{nebmod}%
    \end{figure}

In Fig.\ \ref{nebmod} we represent the model nebula shape that
reproduces our data, as well as the density and velocity distributions
in the extended component. The symmetry axis of this component forms
an angle of $\sim$ 15$^\circ$ with the line of sight, i.e.\ 75 degrees
with respect to the plane of the sky.  
In the extended component, the total density is
assumed to be inversely proportional to the distance to the star, $r$,
with $n$(10$^{16}$cm) = 3 10$^3$ cm$^{-3}$.  The rotational temperature
of the extended nebula decreases linearly with $r$, between 11 K and 5
K. The velocity in this component is radial and its modulus increases
linearly with $r$, between 2.5 \kms\ in the center up to 8 \kms. The CO
relative abundance is assumed to be constant and equal to 3 10$^{-4}$
in all the nebula.  Note that we have tried to keep the model very
simple, in view of the lack of observational information. 

The values of most parameters of the central unresolved component are
uncertain. The velocity field of the compact disk cannot be reliably
deduced from the data, as mentioned above, so only a velocity
dispersion can be estimated for this component. We have taken conditions
for which the $^{12}$CO lines are optically thick, in view of the low
contrast shown in the line core for $^{12}$CO and $^{13}$CO transitions
\citep{bujetal01}. Therefore, $^{12}$CO (and even $^{13}$CO) lines
could be opaque in it, and the derived density and mass in the central
clump are only lower limits.  Assuming high optical depth and the size
limit given above, we can derive a kinetic temperature limit from the
measured brightness, $T_k$ $>$ 60 K.

The total masses of both components are compatible with those derived
in \citet{bujetal01} from $^{13}$CO lines ($\sim$ 10$^{-2}$ \ms\ for
the central component and $\sim$ 3 10$^{-3}$ \ms\ for the extended one;
note that we must correct the mass values in that paper for the
different assumed distance). 

\begin{figure*}
\vspace{-1.5cm}
\hspace{-0.5cm}
\rotatebox{270}{
\resizebox{14cm}{!}{ \includegraphics{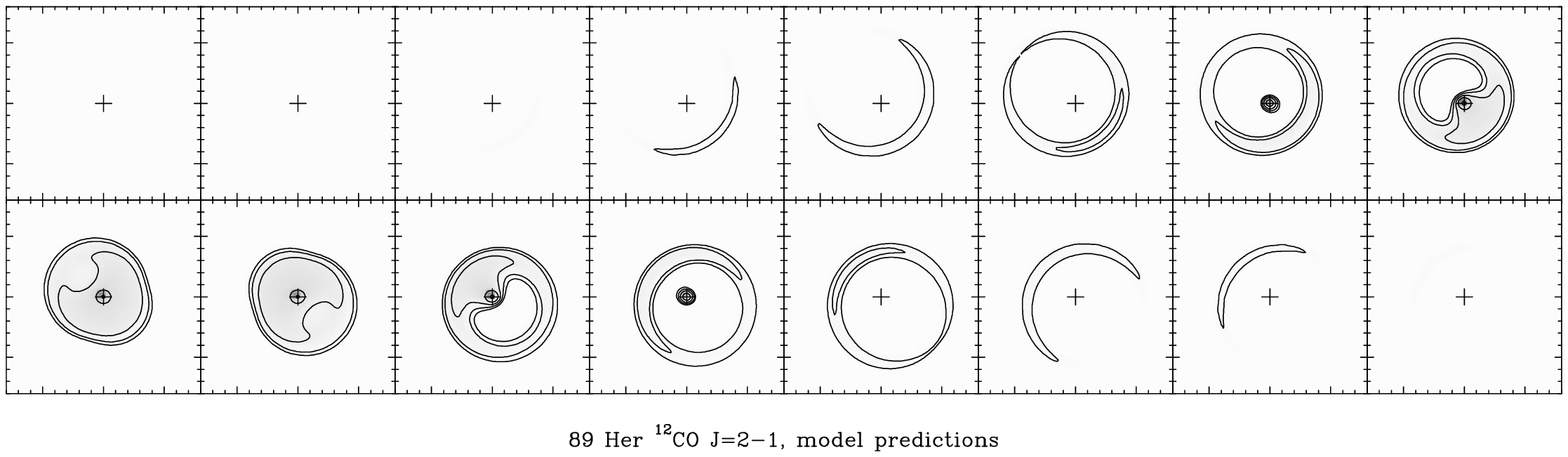}
}}

\vspace{-7.5cm}
   \caption{Synthetic maps predicted by our model of the \doceCO\
            \jdu\ line emission from the nebula around 
            89\,Her. Contours and units are the same as for Fig.\
            \ref{maps}.}
          
              \label{modpre}%
    \end{figure*}

The best fitting of the \jdu\ maps from our model is found in Fig.\
\ref{modpre}. Note that the model necessarily predicts quite symmetric
results for relatively positive and negative velocities, with respect
to the equator, which is not exactly the case in the observations.
Therefore, our fitting can only represent a kind of average of the maps
around the central velocities. The observed asymmetry could correspond
to a small deviation of the symmetry axis for both cups with respect to
the line of sight, by about 10 degrees; a similar deviation for 
the projection of the axis on the plane of the sky is also directly
suggested by our maps.

%

From the size and velocity of the extended component given above, 
we find a typical lifetime $\sim$ 3500 yr. This value is high 
compared to the typical ages of other PPNe, often around 1000
yr, but compatible with expectations for post-AGB stars with
low initial mass \citep[e.g.][]{vassiw94,blocker95}. 

If we also assume that the compact component is in expansion, we can
calculate a typical lifetime. If, as often happens in bipolar nebulae,
it is a disk perpendicular to the symmetry axis of the extended
hour-glass structure (then forming an angle of about 15$^\circ$ with
respect to the plane of the sky), the lifetime must be shorter than
$\sim$ 100 yr. A less strong upper limit is obtained if we assume that
the central structure is spherical and in isotropic expansion, in this
case the lifetime must be shorter than 300 yr.

Note the significant difference between the lifetimes of both extended
and compact components. The very short lifetime derived for the compact
core is surprising. We expect copious mass losses in AGB stars, but not
that an F2-type star ejects about 10$^{-2}$ \ms\ in such a sudden
event, completely independent of the ejection of the extended
component. Moreover, we note that the star is slowly evolving now. It
is known to have its present efficient temperature since at least 1950
\citep[see][]{waters93,bidelman51} up to 2005 (unpublished data), so we
do not expect that it was in the AGB only $\sim$ 100 yr ago.

The very inner nebular structure was also probed by our VLTI 
observations. The total N-band flux is completely
dominated by the circumstellar hot dust component.  Even on our longest
projected baseline (31.1\,m), this N-band emission is barely resolved,
with normalized visibility amplitudes $>$0.7 in the whole spectral band.
The hot dust therefore must come from an angular size smaller than
  $\sim$ 
  31 mas, i.e.\ a diameter \lsim\ 31 AU at the distance
  of 89\,Her.  This result is compatible with the 
  unresolved N-band direct image by  \citet{meixner99}.


If we assume that the central component is a rotating disk, we cannot
derive lifetimes, but we can compare the velocities with the keplerian
ones. Taking a typical projected velocity of $\sim$ 2 \kms, the
characteristic rotation velocity would be $\sim$ 8 \kms\ for
the deduced axis inclination. For a central total mass of 1 \ms\ 
(see mass estimate in Sect.\ 4), 
the rotating gas must
be at a typical distance of about 15 AU, yielding a typical disk size 
$\sim$ 4 10$^{14}$ cm or 30 milli-arcseconds. We note that the outer
diameter of the rotating disk could be larger than these values, 
corresponding to lower rotation velocities but included within the
observed velocity dispersion. This very compact size is strikingly
compatible with  the size measured from our VLT interferometry at 10
$\mu$m. 
We finally note that a typical temperature for the central disk 
$T_k$ $\sim$ 600 K is derived from the measured 
brightness temperature (at relatively high velocities but within the
central component) and a size $\sim$ 0\farcs 05. 
This value is comparable to the temperature of the
warm dust derived from analysis of the SED in 89\,Her and to 
that of the keplerian disk in the Red Rectangle derived from CO
emission modelling, $\sim$ 500 K at 4 10$^{14}$ cm \citep{bujetal05}.

\section{Discussion and conclusions}

Our CO data of 89\,Her show the presence of a double structure, with an
unresolved compact circumstellar component and an hour-glass-like
extended outflow.  Although the velocity field in the inner region is
not well-measured, we conclude that it is probably a keplerian disk
with a very small extent (diameter \lsim\ 10$^{15}$ cm, \lsim\ 0\farcs
1). This is supported by our N-band interferometric data that shows
that, in the N-band, the extent of the dust emission is barely resolved
at a 31.1m baseline. The dust emitting in the 1mm continuum in 89 Her
is also confined to a small central region (diameter $<$ 0\farcs 3),
probably the same compact component.

The existence of rotating inner disks has been proposed to explain some
properties of hot dust emission from other binary post-AGB stars
(Sect.\ 1), properties that are in fact shared by the Red Rectangle and
89 Her. Our results support the identification of inner stable
reservoirs of material in keplerian rotation in these objects.  The
small size of the keplerian disk in 89\,Her can be a common property of
such candidate disks, in view of the high dust temperature measured in
them.  89\,Her and the Red Rectangle are likely the nearest of these
objects \citep{deruyter06}, so that these small regions are not
expected to be resolved by present day radio interferometers, and
a direct confirmation of the keplerian dynamics in the other objects
will require the use of future instruments.  Our N-band interferometric
experiments do confirm the very compact nature of the warm dust
emission in similar objects like HD\,52981 and SX\,Cen \citep{deroo06}.

Assuming that the symmetry axis of the resolved hour-glass outflow is
perpendicular to the inner orbit of the binary, the 
mass function of 0.000838 M$_{\odot}$ \citep{waters93} converts to a
mass of the companion of only 0.35 M$_{\odot}$ (i=10$^{\circ}$ gives
0.62 M$_{\odot}$, i=20$^{\circ}$ yields 0.25 M$_{\odot}$) assuming a
primary of $\sim$ 0.6 M$_{\odot}$. Since there is no evidence of a hot,
compact degenerate component in the system, the companion is likely to
be a very low-mass main sequence dwarf.

With this inclination the orbit has a semi-major axis of 0.31 AU
($a$\,$\sin i$ = 0.08 AU). This orbit is well within the sublimation
radius of the primary, so the circumstellar dusty disk must be
circumbinary. The primary does not fill its Roche-lobe now; but
assuming the star had a similar luminosity when on the AGB, it must
have had a phase of very strong binary interaction.

\begin{acknowledgements}
      J.A. and V.B. acknowledge partial support from the 
      \emph{Spanish Ministry of
     Education \& Science} project numbers AYA2003-7584 and ESP2003-04957
\end{acknowledgements}

\bibliographystyle{aa}
\bibliography{buj}
\end{document}